\documentclass[aps,pre,floatfix,twocolumn,groupedaddress]{revtex4-1}
\usepackage[dvips]{graphicx}
\usepackage{amsmath}
\usepackage{verbatim}
\newcommand{\etalia}{{\it et al.~}}
\newcommand{\la}{\left\langle}
\newcommand{\ra}{\right\rangle}

\newcommand{\PRL}{Phys.~Rev.~Lett.~}
\newcommand{\PR}{Phys.~Rev.~}
\newcommand{\JCP}{J.~Chem.~Phys.~}

\newcommand{\JPCM}{J.~Phys.: Condens.~Matter~}

\begin{document}

\title{Depletion-Induced Forces and Crowding in Polymer-Nanoparticle Mixtures: \\
Role of Polymer Shape Fluctuations and Penetrability}

\author{Wei Kang Lim and Alan R. Denton}
\email[]{alan.denton@ndsu.edu}
\affiliation{Department of Physics, North Dakota State University,
Fargo, ND 58108-6050, USA}

\begin{abstract}
Depletion forces and macromolecular crowding govern the structure and function of 
biopolymers in biological cells and the properties of polymer nanocomposite materials.
To isolate and analyze the influence of polymer shape fluctuations and penetrability 
on depletion-induced interactions and crowding by nanoparticles, we model polymers as 
effective penetrable ellipsoids, whose shapes fluctuate according to the probability 
distributions of the eigenvalues of the gyration tensor of an ideal random walk.
Within this model, we apply Monte Carlo simulation methods to compute 
the depletion-induced potential of mean force between hard nanospheres and crowding-induced 
shape distributions of polymers in the protein limit, in which polymer coils can be
easily penetrated by smaller nanospheres.  By comparing depletion potentials from 
simulations of ellipsoidal and spherical polymer models with predictions of 
polymer field theory and free-volume theory, we show that polymer depletion-induced 
interactions and crowding depend sensitively on polymer shapes and penetrability, 
with important implications for bulk thermodynamic phase behavior.
\end{abstract}

\maketitle
\newpage


\section{Introduction}\label{introduction}
Depletion forces can profoundly influence the properties of soft materials, 
e.g., colloid-polymer mixtures~\cite{lekkerkerker-tuinier2011} and 
polymer nanocomposites~\cite{balazs2006,mackay2006}, 
whose multiple species intermingle and exclude volume to one another.
The exclusion of polymers, or other flexible macromolecules, from a layer
surrounding rigid colloidal particles creates an imbalance in osmotic pressure 
that can induce an effective interparticle pair attraction.  This depletion mechanism, 
recognized by Asakura and Oosawa over 60 years ago~\cite{asakura1954}, is physically 
reasonable and well established in the ``colloid limit", in which the particle radius 
exceeds the average polymer radius of gyration.  In the opposite ``protein limit", 
in which nanoparticles (e.g., globular proteins) can penetrate much larger polymer coils, 
the concept of a depletion layer around a particle must be replaced by that of 
a segment-segment correlation length (blob radius) of the polymer coil~\cite{deGennes1979}.

Depending on the relative strengths of 
competing intermolecular interactions~\cite{israelachvili1992}, including steric, 
electrostatic, and van der Waals interactions, depletion-induced attraction 
can drive aggregation and thermodynamic phase separation, leading, for example,
to coexistence of polymer-rich and 
polymer-poor bulk phases~\cite{vrij1976,pusey1991,jones2002,fuchs2002,fleer-tuinier2008}.
In soft materials, where effective interactions~\cite{denton-zvelindovsky2007} between macromolecules
are typically comparable in magnitude to thermal energies, and self-assembly often depends on 
a delicate balance between energy and entropy, depletion can crucially affect phase stability.
The ability to control depletion forces has great practical value for stabilizing 
foods~\cite{tolstoguzov1991,deKruif-tuinier2001} and pharmaceuticals against coagulation,
purifying water by flocculation and sedimentation of colloidal particles~\cite{norde2011}, 
guiding the self-assembly of virus particles~\cite{dogic2004,li2013}, and
promoting or inhibiting aggregation of proteins~\cite{zukoski2001}, with relevance 
for deciphering protein structure via scattering measurements on crystalline samples 
and treating diseases, such as cataracts~\cite{stradner2007}.

An intrinsic complement to depletion is the phenomenon of macromolecular crowding, 
in which polymers, or other flexible macromolecules, deform in both size and shape 
in response to confinement by other species or bounding surfaces~\cite{minton1981,
minton2000,minton2001,richter2007,richter2008,elcock2010,hancock2012}.
Kuhn's prescient insight~\cite{kuhn1934} that a polymer coil in solution fluctuates in shape, 
being well-approximated by an elongated, flattened ellipsoid (in its principal-axis frame 
of reference), has inspired numerous mathematical and statistical mechanical analyses 
of the shapes of random walks~\cite{fixman1962,flory-fisk1966,flory1969,yamakawa1970,
fujita1970,solc1971,solc1973,theodorou1985,rudnick-gaspari1986,rudnick-gaspari1987,
bishop1988,sciutto1996,schaefer1999,murat-kremer1998,eurich-maass2001}.
In biology, conformational changes of biopolymers, such as RNA, DNA, and unfolded proteins 
are important for cellular processes in the crowded environment of the 
nucleus~\cite{ellis2001a,ellis2001b,vandermaarel2008,
phillips2009,cheung2013,denton-cmb2013},
packaging of DNA in viral capsids~\cite{yeomans2006}, and translocation of biopolymers 
through narrow pores~\cite{polson2013}.

Depletion forces, polymer crowding, and phase behavior in colloid-polymer mixtures 
and polymer-nanoparticle composite materials have been probed by a diverse array of 
experimental methods, including neutron scattering~\cite{tong1996,han2001,
kramer2005a,kramer2005b,kramer2005c,longeville2009,longeville2010,richter2010},
atomic force microscopy~\cite{milling-biggs1995},
total internal reflection microscopy~\cite{leiderer1998}, 
optical trapping~\cite{yodh1998,yodh2001,dogic2015}, and 
turbidity measurements~\cite{vanduijneveldt2005,vanduijneveldt2006,vanduijneveldt2007}.
Related modeling approaches have been based on mean-field and 
scaling theories~\cite{deGennes1979,joanny-leibler-deGennes1979,sear1997,sear2001,sear2002},
free-volume theories~\cite{minton2005,denton-schmidt2002,lu-denton2011,lim-denton2014}, 
force-balance theory~\cite{walz-sharma1994},
perturbation theory~\cite{mao-cates-lekkerkerker1995,mao-cates-lekkerkerker1997},
polymer field theory~\cite{eisenriegler1996,hanke1999,eisenriegler2003,odijk2000,
forsman2010,forsman2012,forsman2014},
integral-equation theory~\cite{chatterjee1998a,schweizer2002,moncho-jorda2003},
density-functional theory~\cite{leiderer1999,schmidt-fuchs2002,mukherjee2004,
forsman2008,forsman2009}, 
adsorption theory~\cite{tuinier-lekkerkerker2000,tuinier-lekkerkerker2001,tuinier-petukhov2002},
and simulation of both molecular~\cite{meijer-frenkel1991,meijer-frenkel1994,dickman1994,
bolhuis2002,louis2002-jcp2,bolhuis2003,doxastakis2004,doxastakis2005,goldenberg2003,dima2004,cheung2005,
denesyuk2011,denesyuk-thirumalai2013a,likos2010,linhananta2012,wittung-stafshede2012,cheung2013}
and coarse-grained~\cite{hoppe2011,denesyuk-thirumalai2013b,lu-denton2011,lim-denton2014} polymer models.

Previous studies have explored the nature of depletion interactions induced by 
aspherical depletants of fixed size and shape~\cite{mao-cates-lekkerkerker1997,kamien1999,
piech-walz2000,yodh2001} and the impact of crowding on polymer size~\cite{kramer2005a,
kramer2005b,kramer2005c,longeville2009,longeville2010,richter2010,minton2005,goldenberg2003,
dima2004,cheung2005,denesyuk2011,denesyuk-thirumalai2013a}.
In recent work, we reported on preliminary studies of polymer crowding in models 
of polymer-nanoparticle mixtures with polymers modeled as fluctuating, penetrable 
spheres~\cite{lu-denton2011} or ellipsoids~\cite{lim-denton2014}.  In this paper, 
using a refined model of polymer-nanoparticle penetration, we directly analyze 
the complex relationships between polymer depletion-induced interactions, 
nanoparticle crowding, and polymer shape fluctuations. 
By comparing results with theoretical predictions, we validate the polymer model and
demonstrate the significance of shape fluctuations in depletion and crowding
phenomena.  Our model and computational methods are sufficiently general
as to be easily adapted and applied to other soft materials.

The remainder of the paper is assembled as follows.  In Sec.~\ref{models}, we review 
the model of polymers as fluctuating penetrable ellipsoids and the field theory model  
for the free energy cost of penetrating a polymer by a hard nanosphere.
In Sec.~\ref{methods}, we outline our simulation methods for computing polymer 
depletion-induced interactions between nanospheres and crowding-induced deformations 
in polymer shape.  
Numerical results are presented in Sec.~\ref{methods} and compared with predictions of 
field theories and free-volume theory.
Finally, in Sec.~\ref{methods}, we conclude with suggestions for future work.


\section{Models}\label{models}
\subsection{Ellipsoidal Polymer Model}\label{ellipsoid-model}

\begin{figure}
\begin{center}
\includegraphics[width=0.9\columnwidth]{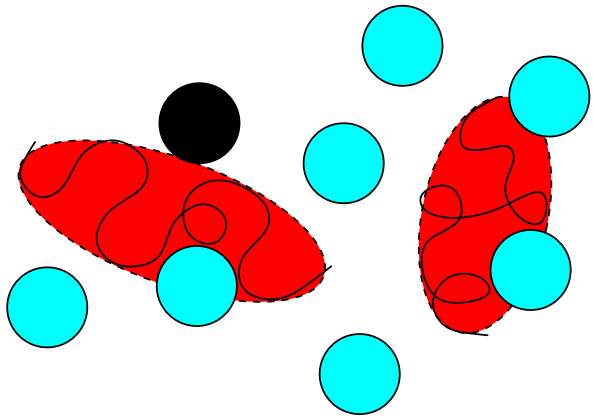}
\end{center}
\caption{Model of a polymer-nanoparticle mixture with polymers 
represented as ellipsoids that can fluctuate in size and shape.  
Nanoparticles are hard spheres of fixed size that are mutually impenetrable,
but able to penetrate the polymers.  Penetration is defined as the
intersection of the surfaces of an ellipsoid and a sphere.  For example, 
the dark nanosphere is just on the verge of penetrating a polymer.
}\label{fig-model}
\end{figure}

Our model extends the classic Asakura-Oosawa-Vrij (AOV) model of colloid-polymer 
mixtures~\cite{asakura1954,vrij1976}.  The AOV model represents nonadsorbing polymer coils, 
in a coarse-grained approximation, as effective spheres of fixed size that are 
mutually noninteracting, but impenetrable to the colloidal particles.  
The spherical polymer approximation ignores fluctuations in conformation,
while the impenetrable polymer assumption is justified only in the colloid limit,
where the polymers are smaller than the particles.
As in our previous studies~\cite{lim-denton2014,lim-denton2015}, we refine
the AOV model by representing the polymers as soft ellipsoids that fluctuate 
in size and shape and that can be penetrated by smaller nanoparticles. 

For simplicity, we focus on linear homopolymers, although the analysis can be
generalized to other polymer architectures, such as block copolymers~\cite{eurich2007}.
A coil of $N$ identical, connected segments has a size and shape characterized by 
the gyration tensor ${\bf T}$ with components
\begin{equation}
T_{ij}=\frac{1}{N}\sum_{k=1}^N r_{ki} r_{kj}~, 
\label{gyration-tensor}
\end{equation}
where $r_{ki}$ is $i^{\rm th}$ component of the position vector ${\bf r}_k$ 
of the $k^{\rm th}$ segment relative to the center of mass.
The familiar moment of inertia tensor ${\bf I}$ of rigid body dynamics relates to the 
gyration tensor via ${\bf I}=R_p^2{\bf 1}-{\bf T}$, where ${\bf 1}$ is the unit tensor and
\begin{equation}
R_p=\left(\frac{1}{N}\sum_{i=1}^Nr_i^2\right)^{1/2}=(\Lambda_1+\Lambda_2+\Lambda_3)^{1/2}
\label{radius-gyration}
\end{equation}
is the radius of gyration of a given conformation of the coil expressed in terms of
the eigenvalues $\Lambda_i$ ($i=1,2,3$ in three dimensions) of ${\bf T}$.
The experimentally measurable root-mean-square (rms) radius of gyration is given by
\begin{equation}
R_g=\sqrt{\la R_p^2\ra}=\sqrt{\la\Lambda_1+\Lambda_2+\Lambda_3\ra}~,
\label{gyration-avg}
\end{equation}
where the angular brackets represent an ensemble average over polymer conformations.

If the ensemble average in Eq.~(\ref{gyration-avg}) is evaluated in a reference frame  
tied to the principal axes of the coil,
and the coordinate axes are labelled to preserve the order of the eigenvalues by
magnitude ($\Lambda_1>\Lambda_2>\Lambda_3$), then the average tensor is asymmetric
and describes an anisotropic object~\cite{rudnick-gaspari1986,rudnick-gaspari1987}.
Averaging in a fixed (laboratory) frame yields, in contrast, a symmetric average tensor 
that has equal eigenvalues and thus describes a sphere.  
Simply stated, a fluctuating random walk has an average shape that is spherical 
when viewed from the laboratory frame, but significantly aspherical -- elongated, flattened, 
bean-shaped -- when viewed from the principal-axis frame~\cite{kuhn1934,solc1971,solc1973}.
The general ellipsoid that best fits the shape of the polymer coil -- roughly corresponding
to the tertiary structure of a biopolymer -- has principal radii proportional to the 
square-roots of the respective eigenvalues of the gyration tensor. 
In Cartesian $(x,y,z)$ coordinates, the surface is described by
\begin{equation}
\frac{x^2}{\Lambda_1}+\frac{y^2}{\Lambda_2}+\frac{z^2}{\Lambda_3}=3~.
\label{ellipsoid}
\end{equation}

The probability distribution for the shape of a freely-jointed polymer coil 
of $N$ segments of length $l$, modeled as a soft ellipsoid~\cite{murat-kremer1998}, 
is accurately approximated by the analytical form~\cite{eurich-maass2001}
\begin{equation}
P_0(\lambda_1,\lambda_2,\lambda_3) = P_1(\lambda_1)P_2(\lambda_2)P_3(\lambda_3)~,
\label{Plambda}
\end{equation}
where $\lambda_i\equiv\Lambda_i/(Nl^2)$ are scaled (dimensionless) eigenvalues and 
\begin{equation}
P_i(\lambda_i) = \frac{(a_id_i)^{n_i-1}\lambda_i^{-n_i}}{2K_i}
\exp\left(-\frac{\lambda_i}{a_i}-d_i^2\frac{a_i}{\lambda_i}\right)~,
\label{Eurich-Maass}
\end{equation}
with fitting parameters 
$K_1=0.094551$, $K_2=0.0144146$, $K_3=0.0052767$, 
$a_1=0.08065$, $a_2=0.01813$, $a_3=0.006031$,
$d_1=1.096$, $d_2=1.998$, $d_3=2.684$, 
$n_1=1/2$, $n_2=5/2$, and $n_3=4$. 
Although the factorization ansatz of Eq.~(\ref{Plambda}) is not exact, since extensions
of a random walk in orthogonal directions are not independent, conformations that 
seriously violate the ansatz are rare for sufficiently long random walks.
It is important to note that the eigenvalue distributions [Eq.~(\ref{Eurich-Maass})] are 
derived from random-walk segment statistics~\cite{murat-kremer1998,eurich-maass2001},
reflect considerable fluctuations in size and shape of a free polymer, and will be 
modified by confinement, e.g., for polymers in the presence of nanoparticle crowders.

\subsection{Polymer-Nanoparticle Mixtures}\label{mixtures-model}
A mixture of hard nanospheres and nonadsorbing polymers is characterized by the 
number densities of the two species, $n_n$ and $n_p$, the nanosphere radius $R_n$, and the 
rms radius of gyration $R_g$ of free (uncrowded) polymer.
In the colloid limit ($R_g<R_n$), in which the polymer coils are impenetrable to particles,
the effective polymer size must be calibrated in order to consistently and accurately account 
for the polymer excluded volume~\cite{lim-denton2015}.
In the protein limit ($R_g\gg R_n$), in which polymer penetration supplants excluded volume,
it is instead the penetration energy that must be calibrated, as explained below 
(Sec.~\ref{penetrable-model}).  Thus, we simply take the polymer radius equal to $R_g$ and
define the polymer-to-nanosphere size ratio as $q\equiv R_g/R_n$.  The volume fractions 
of the two species are $\phi_n\equiv (4\pi/3)n_n R_n^3$ and $\phi_p\equiv (4\pi/3)n_p R_g^3$.
It should be noted that the actual fraction of volume occupied by polymer coils will differ 
from $\phi_p$ due to the aspherical shapes and interpenetration of the coils.

In terms of the scaled eigenvalues, 
deviations of a polymer's average shape from spherical can be quantified
by an asphericity parameter~\cite{rudnick-gaspari1986,rudnick-gaspari1987}
\begin{equation}
A(\phi_n)=1-3\frac{\la\lambda_1\lambda_2+\lambda_1\lambda_3+\lambda_2\lambda_3\ra}
{\la(\lambda_1+\lambda_2+\lambda_3)^2\ra}~.
\label{asphericity}
\end{equation}
A spherical object with all three eigenvalues equal has $A=0$, while a greatly elongated object, 
with one eigenvalue much larger than the other two, has $A\simeq 1$.
The ratio of the rms radius of gyration [Eq.~(\ref{gyration-avg})] of a polymer coil 
crowded by nanoparticles, $R_g(\phi_n)$, 
to that of an uncrowded coil, $R_g(0)=l\sqrt{N/6}$, can be expressed as
\begin{equation}
\frac{R_g(\phi_n)}{R_g(0)}=\sqrt{6\la\lambda_1+\lambda_2+\lambda_3\ra}~,
\label{gyration-ratio}
\end{equation}
while the principal radii of the representative ellipsoid are given by 
\begin{equation}
R_i(\phi_n)=R_g(0)\sqrt{18\lambda_i}~, \quad i=1,2,3~.
\label{principal-radii}
\end{equation}

\subsection{Polymer-Nanoparticle Penetration}\label{penetrable-model}
In the protein limit ($q\gg 1$), a nanoparticle may penetrate a polymer coil, 
with a free energy cost associated with the reduction in conformational entropy of the coil. 
The average free energy cost $f$ to insert a hard sphere into an ideal polymer solution  
at temperature $T$ is predicted by 
polymer field theory~\cite{eisenriegler1996,hanke1999,eisenriegler2003}:
\begin{equation}
f=k_B T~\frac{4\pi n_p R_p^3}{q}\left(1+\frac{2}{\sqrt{\pi}q}+\frac{1}{3q^2}\right)~,
\label{f}
\end{equation}
which is valid for all $q$ (Eq.~(3.11) of ref.~\cite{eisenriegler1996}).

The simplest model of the pair interaction between a polymer and a nanoparticle,
proposed by Schmidt and Fuchs~\cite{schmidt-fuchs2002} in modeling phase behavior of
polymer-nanoparticle mixtures, treats the penetration energy profile as a step function, 
equal to a constant $\varepsilon$ in the case of penetration and zero otherwise.  
For a polymer of average volume $v_p$, this model predicts an average insertion free energy 
$n_pv_p\varepsilon$.  Equating $n_pv_p\varepsilon$ to $f$ in Eq.~(\ref{f}) yields 
\begin{equation}
\beta\varepsilon=
\frac{4\pi R_p^3}{v_p q}\left(1+\frac{2}{\sqrt{\pi}q}+\frac{1}{3q^2}\right)~,
\label{epsilon}
\end{equation}
where $\beta\equiv 1/(k_B T)$.  If the polymer were approximated as a sphere 
of radius $R_p$, then Eq.~(\ref{epsilon}) would yield 
\begin{equation}
\beta\varepsilon=\frac{3}{q}\left(1+\frac{2}{\sqrt{\pi}q}+\frac{1}{3q^2}\right)~,
\label{epsilon-sphere}
\end{equation}
or $\beta\varepsilon\simeq 3/q$ for $q\gg 1$.  
Schmidt and Fuchs~\cite{schmidt-fuchs2002} applied the latter approximation in their
study of demixing, as did we in our previous studies of crowding~\cite{lu-denton2011,lim-denton2014}.
Here we apply, however, a more consistent and accurate calibration, which evaluates 
the average volume of the ellipsoidal polymer from the true shape distribution.
For an uncrowded polymer, Eqs.~(\ref{Plambda}) and (\ref{Eurich-Maass}) yield
\begin{equation}
v_p = \frac{4\pi}{3}\int d\lambda\, P_0(\lambda) R_1R_2R_3 = 1.8365~R_p^3~,
\label{volume-ellipsoid0}
\end{equation}
with $\lambda\equiv(\lambda_1,\lambda_2,\lambda_3)$.
Substituting this polymer volume into Eq.~(\ref{epsilon}) then leads to
\begin{equation}
\beta\varepsilon=\frac{6.8426}{q}\left(1+\frac{2}{\sqrt{\pi}q}+\frac{1}{3q^2}\right)~.
\label{epsilon-ellipsoid}
\end{equation}
For a crowded polymer, amidst nanoparticles of volume fraction $\phi_n$, 
whose shape distribution we denote as $P(\lambda; \phi_n)$
with factors $P_i(\lambda_i; \phi_n)$, 
Eq.~(\ref{volume-ellipsoid0}) must be modified accordingly:
\begin{equation}
v_p(\phi_n) = \frac{4\pi}{3}\int d\lambda\, P(\lambda; \phi_n) R_1R_2R_3~.
\label{volume-ellipsoid}
\end{equation}
In the simulations described below, we computed the polymer-nanosphere penetration 
free energy from Eq.~(\ref{epsilon}), with $v_p$ consistently determined from 
Eq.~(\ref{volume-ellipsoid}).

\section{Methods}\label{methods}
\subsection{Monte Carlo Simulations}\label{simulation}
For the models described in Sec.~\ref{models}, we developed simulation methods 
to compute the potential of mean force (PMF) between hard nanospheres induced by 
larger polymers that fluctuate in shape according to Eq.~(\ref{Eurich-Maass}) and 
the shape distributions of crowded polymers immersed in nanosphere dispersions.
The present analysis significantly extends our previous work in which we computed
the PMF in the colloid limit~\cite{lim-denton2015} and crowding effects in the 
protein limit with a cruder penetration model~\cite{lim-denton2014}.

We first summarize the Metropolis Monte Carlo (MC) simulation algorithm 
in the canonical ensemble.  In a simulation cell shaped as either a cube or a 
rectangular parallelepiped (right rectangular prism) of fixed volume with 
periodic boundary conditions, containing a fixed number of particles at constant 
temperature, we performed trial moves comprising displacements of hard nanospheres 
and displacements, rotations, and shape deformations of penetrable ellipsoidal polymers.
With the exception of polymer shape changes, the trial moves were accepted with 
probability~\cite{frenkel-smit2001,binder-heermann2010}
\begin{equation}
{\cal P}_{\rm config}({\rm old}\to{\rm new})=
\min\left\{e^{-\beta\Delta F},~1\right\}~,
\label{disp-rot}
\end{equation}
where $\Delta F$ is the associated change in free energy.  While nanosphere-nanosphere overlaps 
are rejected outright, polymer-nanosphere overlaps are accepted with probability 
${\cal P}_{\rm config}$ using the penetration free energy of Eq.~(\ref{epsilon-ellipsoid}).
For a move that creates/eliminates an overlap, $\Delta F=\pm\varepsilon$.
Intersection of polymer-nanosphere pairs is diagnosed using an essentially exact 
overlap algorithm that computes the shortest distance between the surfaces of a sphere 
and an ellipsoid~\cite{heckbert1994}.
Defining the orientation of a polymer coil by a unit vector ${\bf u}$, aligned with 
the long axis of the ellipsoid, trial rotations are executed by generating a new (trial) 
direction ${\bf u}'$ via
\begin{equation}
{\bf u}'=\frac{{\bf u}+\tau{\bf v}}{|{\bf u}+\tau{\bf v}|}~,
\label{rotation}
\end{equation}
where ${\bf v}$ is a randomly oriented unit vector and the  tolerance $\tau$ 
determines the magnitude of the rotation~\cite{frenkel-smit2001}.
A trial change in shape of an ellipsoidal polymer coil from gyration tensor eigenvalues 
$\lambda$ to new eigenvalues $\lambda'$ 
is accepted with probability 
\begin{equation}
{\cal P}_{\rm shape}(\lambda\to\lambda') = 
\min\left\{\frac{P_0(\lambda')}{P_0(\lambda)}
e^{-\beta\Delta F},~1\right\}~,
\label{shape-variation}
\end{equation}
where $P_0(\lambda)$ is the shape distribution of the uncrowded polymer  
[Eqs.~(\ref{Plambda}) and (\ref{Eurich-Maass}) for ideal polymers].
Trial changes in eigenvalues allow the polymers to evolve toward a new equilibrium 
shape distribution, modified by the presence of nanosphere crowders.
Although we focus here on ideal polymers, it is important to note that our simulation 
method can be easily extended to nonideal polymers with excluded-volume interactions 
by substituting the appropriate shape distribution into Eq.~(\ref{shape-variation}).

\subsection{Potential of Mean Force Algorithm}\label{pmf}

For two nanospheres in thermal and chemical equilibrium with a reservoir of nonadsorbing 
polymers of bulk density $n_p$ at constant $T$, the potential of mean force (PMF) is 
defined as the change in grand potential $\Omega(r)$ upon bringing the nanospheres from 
infinite to finite (center-to-center) separation $r$:
\begin{equation}
v_{\rm mf}(r)=\Omega(r)-\Omega(\infty)~,
\label{veff1}
\end{equation}
where we use the fact that in an isotropic fluid the pair potential depends on 
only the radial coordinate.  The change in grand potential arises from 
mechanical ($pV$) work performed by the nanoparticles in pushing against the 
osmotic pressure of the polymers: $\Pi_p=n_p k_BT$ for ideal polymers.
In the spherical polymer (AOV) model, this work is easily evaluated:
\begin{equation}
v_{\rm mf}(r)=-\Pi_p\int_{\infty}^r dr'\, A_{\rm ov}(r')=-\Pi_p V_{\rm ov}(r)~,
\label{veff2}
\end{equation}
where $A_{\rm ov}(r)$ and $V_{\rm ov}(r)$ are the cross-sectional area and volume, 
respectively, of the overlap region of the two excluded-volume shells
and we choose $\Omega(\infty)=0$.
The convex-lens-shaped overlap region has volume 
\begin{equation}
V_{\rm ov}(r)=
\frac{\displaystyle 4\pi}{\displaystyle 3}
\left[\left(R_n+R_p\right)^3
-\frac{\displaystyle 3r}{\displaystyle 4}(R_n+R_p)^2+
\frac{\displaystyle r^3}{\displaystyle 16}\right]~, 
\label{AOV}
\end{equation}
for $2R_n<r<2(R_n+R_p)$ (zero otherwise).  This simple geometric approach fails, however,
for the fluctuating ellipsoidal polymer model, for which calculating $V_{\rm ov}(r)$ 
requires averaging over polymer shapes and orientations.  As an alternative approach, 
we derive below a more general expression for the PMF, which we evaluate numerically via 
MC simulation using a variation of the Widom particle insertion method~\cite{widom1963}. 

The particle insertion method exploits the connection between the grand potential
$\Xi$ and the grand canonical partition function: $\Omega=-k_B T\ln \Xi$.
For a polymer solution containing two nanospheres -- one fixed at the origin,
the other fixed at a distance $r$ from the origin -- the partition function
is proportional to a configurational integral of the Boltzmann factor for the 
internal potential energy:
\begin{equation}
\Xi(r)\propto \la\exp[-\beta U(r)]\ra~,
\label{Xi}
\end{equation}
where $U(r)$ is the potential energy of the system with two nanospheres at separation $r$
and $\la~\ra$ still represents an ensemble average over polymer conformations.
From Eqs.~(\ref{veff1}) and (\ref{Xi}), 
\begin{equation}
\beta v_{\rm mf}(r)=-\ln\left(\frac{\la\exp[-\beta U(r)]\ra}{\la\exp[-\beta U(\infty)]\ra}\right)~,
\label{Widom2}
\end{equation}
which, in the dilute limit ($\phi_n\to 0$), wherein $v_{\rm mf}(r)$
becomes simply proportional to $\phi_n$, reduces to
\begin{equation}
\beta v_{\rm mf}(r)=\la\exp[-\beta U(\infty)]\ra - \la\exp[-\beta U(r)]\ra~.
\label{Widom3}
\end{equation}

In practice, we computed the PMF induced by ideal polymers as follows.  Fixing a single 
nanoparticle at the origin, we inserted a polymer of random shape and orientation at a 
random position in the box, computed the resultant overlap potential energy $U_0$,
and accumulated the average of $\exp(-\beta U_0)$ over many insertions.
This average, equal to the mean free-volume fraction of the polymer, can be expressed 
as $\la\exp(-\beta U_0)\ra=1-c(r)\phi_n$, where $c(r)$ is independent of $\phi_n$.
It follows that
\begin{equation}
\la\exp[-\beta U(\infty)]\ra=1-2c(r)\phi_n=2\la\exp(-\beta U_0)\ra-1~.
\label{Uinfinity1}
\end{equation}
Note that, because ideal polymers are independent, we need only insert a single polymer at a time.
Repeating for two nanoparticles fixed at separation $r$, we calculated the overlap potential 
energy $U(r)$ upon insertion of a polymer and accumulated the average of $\exp[-\beta U(r)]$ 
over many insertions to obtain $\la\exp[-\beta U(r)]\ra$ and finally $v_{\rm mf}(r)$ 
from Eqs.~(\ref{Widom3}) and (\ref{Uinfinity1}).

As a consistency check, in the AOV model, the PMF can be derived exactly.
In the presence of a single nanoparticle, the average fraction 
of the total volume $V$ available to a polymer is
\begin{equation}
\la\exp(-\beta U_0)\ra
=\frac{V-\frac{4\pi}{3}(R_n+R_p)^3}{V}
=1-\phi_n(1+q)^3~.
\label{alpha0}
\end{equation}
Substituting into Eq.~(\ref{Uinfinity1}) yields
\begin{equation}
\la\exp[-\beta U(\infty)]\ra=1-2\phi_n(1+q)^3~.
\label{Uinfinity2}
\end{equation}
In the presence of two nanoparticles, the average polymer free-volume fraction is
\begin{equation}
\la\exp[-\beta U(r)]\ra
=\frac{V-\frac{8\pi}{3}(R_n+R_p)^3+V_{\rm ov}(r)}{V}~.
\label{alpha1}
\end{equation}
Substituting for $V_{\rm ov}(r)$ from Eq.~(\ref{AOV}) then yields
\begin{equation}
\la\exp[-\beta U(r)]\ra=
1-\phi_n\left[(1+q)^3+\frac{3}{4}x(1+q)^2-\frac{x^3}{16}\right]~,
\label{alpha2}
\end{equation}
with $x\equiv r/\sigma_n$.  The difference of Eqs.~(\ref{alpha2}) and (\ref{Uinfinity2}) 
yields finally the AOV expression for the PMF induced by a single polymer 
[Eqs.~(\ref{veff2} and (\ref{AOV})].
Comparing analytical and simulation results for the AOV model
provides a test of our numerical algorithm (see Sec.~\ref{results}).

\subsection{Theoretical Approaches}\label{theory}
We now summarize theoretical approaches with whose predictions we compare
our simulation results in Sec.~\ref{results}.  The polymer-induced potential of mean force 
between nanoparticles is predicted by polymer field theories.
Within a continuum chain model of monodisperse homopolymers,
Eisenriegler \etalia\cite{eisenriegler1996,hanke1999,eisenriegler2003}
solved a diffusion equation for the partition function of an ideal
(non-self-avoiding) chain in the presence of impenetrable mesoscopic particles,
thus obtaining the first terms in a small-particle ($1/q$) series expansion 
for the average free energy of immersing both a single hard nanosphere [Eq.~(\ref{f})] 
and a pair of nanospheres in an ideal polymer solution.  Including the leading 
and next-to-leading contributions, which should suffice in the protein limit 
($q\gg 1$), their result for the PMF can be expressed as
(see Eqs.~(2.64) and (2.65) of ref.~\cite{eisenriegler2003})
\begin{equation}
\beta v_{\rm mf}(r)=-12\phi_p\left[\frac{h(x/q)}{qx}
+\frac{g(x/q)}{q^2x} - \frac{h(2x/q)}{2qx^2}\right]~,
\label{Eisenriegler}
\end{equation}
where 
\begin{equation}
g(x)\equiv\frac{e^{-x^2}}{\sqrt{\pi}}-x+x{\rm Erf}(x)
\label{g}
\end{equation}
and
\begin{equation}
h(x)\equiv\frac{1}{4}\left[-\frac{2}{\sqrt{\pi}}x e^{-x^2}+(1+2x^2){\rm Erfc}(x)\right]~,
\label{h}
\end{equation}
with ${\rm Erf(x)}$ and ${\rm Erfc(x)}$ being the error function and 
complementary error function, respectively.

In related work, Woodward and Forsman~\cite{forsman2010,forsman2012} and 
Wang \etalia\cite{forsman2014} developed a general field theory, also within the
continuum chain model, for the interaction between nanospheres immersed in 
a fluid of polydisperse ideal homopolymers.  By solving a Schr\"odinger-like equation 
for the end-end segment distribution function, these workers derived an exact expansion 
in spherical harmonics for the PMF induced by polymers with molecular weight 
following the Schulz-Flory distribution:
\begin{equation}
p^{(n)}(s)=\frac{(n+1)^{n+1}}{\Gamma(n+1)}\frac{s^n}{{\bar s}^{n+1}}
\exp[-(n+1)s/{\bar s}]~,
\label{SF}
\end{equation}
with $s$ the degree of polymerization and ${\bar s}$ its mean value.
Their prediction for the PMF reduces to that of Eisenriegler \etalia\cite{eisenriegler2003} 
in the monodisperse ($n\to\infty$) limit 
and readily yields to numerical solution in the case $n=0$.
Although several other approaches to modeling depletion potentials 
have been proposed (cited in Sec.~\ref{introduction}),
we focus here on polymer field theories, since their predictions can be 
directly compared with our simulation results.

To model polymer crowding, we recently developed a free-volume theory,
which we applied to calculate polymer shape distributions in polymer-nanosphere
mixtures~\cite{denton-cmb2013,lim-denton2014}.  In this approach, the shape distribution 
of crowded polymers is given by
\begin{equation}
P(\lambda;\phi_n) = P_0(\lambda)\frac{\alpha(\lambda;\phi_n)}{\alpha_{\rm eff}(\phi_n)}~,
\label{Plambda-fvt}
\end{equation}
where $\alpha(\lambda; \phi_n)$ is the free-volume fraction of a polymer of
shape $\lambda$ amidst nanoparticles of volume fraction $\phi_n$ and
\begin{equation}
\alpha_{\rm eff}(\phi_n) \equiv \int_0^{\infty}{\rm d}\lambda\, P_0(\lambda)\alpha(\lambda;\phi_n)
\label{alphaeff-fvt}
\end{equation}
is an effective polymer free-volume fraction, 
defined as an average of $\alpha(\lambda; \phi_n)$ over uncrowded polymer shapes.
In the dilute nanoparticle concentration limit ($\phi_n\to 0$), we have $\alpha(\lambda; 0)=1$ 
and the shape distribution reduces to that of the uncrowded polymer:
$P(\lambda; 0)=P_0(\lambda)$.

The essential input to the theory is the polymer free-volume fraction, 
which is well approximated by the 
geometry-based theory of Oversteegen and Roth~\cite{oversteegen-roth2005}.  
By separating thermodynamic properties of the nanosphere crowders from
geometric properties of the polymer depletants, 
using fundamental-measures density-functional theory~\cite{rosenfeld1989,rosenfeld1997,schmidt2000},
this approach generalizes scaled-particle theory~\cite{lebowitz1964} from spheres 
to arbitrary shapes, yielding
\begin{equation}
\alpha(\lambda;\phi_n) = (1-\phi_n')\exp[-\beta(pv_p+\gamma a_p+\kappa c_p)]~,
\label{alpha-fmt}
\end{equation}
where $\phi_n'\equiv\phi_n(1-e^{-\beta\varepsilon})$
and $p$, $\gamma$, and $\kappa$ are the bulk pressure, surface tension at a 
planar hard wall, and bending rigidity of the nanoparticles, while $v_p$, $a_p$, 
and $c_p$ are the volume, surface area, and integrated mean curvature of a polymer.
For a general ellipsoidal polymer, $v_p=(4\pi/3)R_1R_2R_3$, with the principal-radii 
$R_i$ given by Eq.~(\ref{principal-radii}), while $a_p$ and $c_p$ 
are numerically evaluated from the principal radii.  The thermodynamic properties 
of hard nanospheres are accurately approximated by the Carnahan-Starling 
expressions~\cite{oversteegen-roth2005,hansen-mcdonald2006}:
\begin{eqnarray}
\beta p&=&\frac{3\phi_n}{4\pi R_n^3}\frac{1+\phi_n+\phi_n^2-\phi_n^3}{(1-\phi_n)^3}
\nonumber \\[0.5ex]
\beta\gamma&=&\frac{3}{4\pi R_n^2}\left[\frac{\phi_n(2-\phi_n)}{(1-\phi_n)^2}
+\ln(1-\phi_n)\right] \nonumber \\[0.5ex]
\beta\kappa&=&\frac{3\phi_n}{R_n(1-\phi_n)}~.
\label{CSthermo}
\end{eqnarray}

\section{Results and Discussion}\label{results}
To test the accuracy of the ellipsoidal polymer model in describing
polymer depletion-induced interactions between nanoparticles and nanoparticle-induced
crowding of polymers, as well as to validate our MC algorithms, we performed 
two series of simulations in the protein limit.  In one series, we computed the PMF 
between pairs of hard nanospheres; in the second series, we computed the shape distributions 
of polymers crowded by many nanospheres.  
In this section, we compare our results for the depletion potential and polymer crowding 
with predictions of polymer field theory and free-volume theory, respectively.  

In the first series of simulations, we computed the PMF over a range of nanosphere 
separations, using the polymer trial insertion method outlined in Sec.~\ref{pmf}.
The dimensions of the rectangular parallelepiped simulation cell 
were set to maximize the acceptance ratio, while avoiding interaction of polymers 
with periodic images of the nanospheres.
Tolerances for polymer trial moves were fixed at $\tau=0.001$ for rotations 
and $\Delta\lambda_1=0.01$, $\Delta\lambda_2=0.003$, $\Delta\lambda_3=0.001$ for shape changes.
Each run comprised $2\times 10^7$ independent polymer insertions.  
We determined statistical uncertainties (error bars) by computing 
standard deviations from five independent runs.
As a test, we first simulated the original AOV model of spherical polymers, fixed in size 
and impenetrable to nanospheres, and confirmed that our algorithm reproduces the 
exact PMF of Eqs.~(\ref{veff2}) and (\ref{AOV}) to within statistical uncertainties.  
Next, we simulated the penetrable ellipsoidal polymer model (Sec.~\ref{penetrable-model}),
with penetration free energy given by Eq.~(\ref{epsilon-ellipsoid}).  For comparison,
we also simulated a modified AOV model of spherical polymers, fixed in size but penetrable 
to the nanospheres, with the penetration free energy given by Eq.~(\ref{epsilon-sphere}).

\begin{figure}
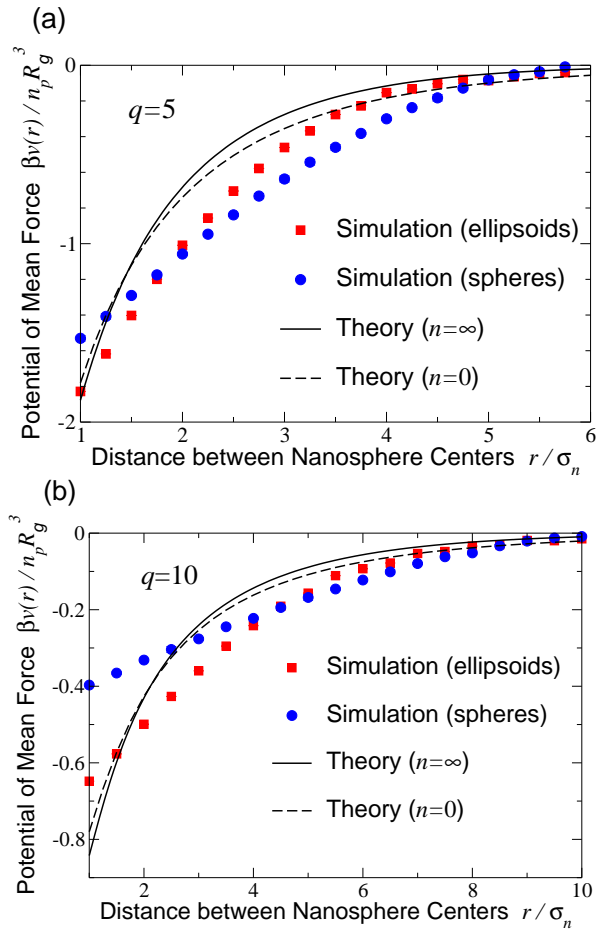

\includegraphics[width=0.9\columnwidth]{fig2a.eps} 
\includegraphics[width=0.9\columnwidth]{fig2b.eps} 
\vspace*{-0.2cm}
\caption{
Potential of mean force (units of $n_pR_g^3k_BT$) induced by ideal polymers between 
hard nanospheres for polymer-to-nanosphere size ratio (a) $q=5$ and (b) $q=10$.  
Our simulation data for ellipsoidal polymers (squares) and spherical polymers (circles) 
are compared with predictions of polymer field theory for monodisperse 
coils~\cite{eisenriegler2003} ($n=\infty$, solid curves) and polydisperse 
coils~\cite{forsman2014} ($n=0$, dashed curves).
Error bars are smaller than symbol sizes.
}\label{fig-pmf-theory}
\end{figure}

Figure~\ref{fig-pmf-theory} presents direct comparisons of our simulation data with
predictions of two polymer field theories~\cite{eisenriegler2003,forsman2014} 
for size ratios $q=5$ and $q=10$.  The two theories make slightly different predictions, 
since one~\cite{eisenriegler2003} describes monodisperse coils [$n=\infty$ in Eq.~(\ref{SF})] 
and the other~\cite{forsman2014} polydisperse coils ($n=0$).
(When replotted in Fig.~\ref{fig-pmf-theory}, the pair potentials from ref.~\cite{forsman2014}
have been corrected for a missing scale factor~\cite{ForsmanNote}.)
With no fitting parameters, the ellipsoidal polymer model yields a PMF in good
agreement with field theories, although with somewhat less curvature.  In contrast, 
the penetrable spherical polymer model, with penetration free energy given by 
Eq.~(\ref{epsilon-sphere}), produces a PMF with a shallower attractive well and 
a qualitatively different shape.
For reference, the AOV model of impenetrable spherical polymer greatly overestimates 
the depth of the attraction, yielding contact values of $\beta v_{\rm mf}(\sigma_n)/(n_pR_g^3)=5.4$
for $q=5$ and 4.8 for $q=10$ -- hardly surprising in the protein limit, where polymers 
are far from impenetrable.  
Quantitative discrepancies between simulation and theory, especially at intermediate
distances, may originate from the step-function approximation for the 
polymer-nanoparticle penetration energy profile. 

Evidently, polymer shape fluctuations {\it and} penetrability both play vital roles 
in depletion-induced interactions.  
In passing, we note that using $\beta\varepsilon=3/q$ as the penetration energy 
for the ellipsoidal polymer model leads to a much weaker pair attraction.
Close agreement between polymer field theory predictions and the potentials output by our simulations, 
which use as input the penetration free energy predicted by the same theory, not only
validates the ellipsoidal polymer model, but also confirms the self-consistency 
of the field theories.  We emphasize, however, that our approach, unlike the field theories, 
can be applied also in the colloid limit~\cite{lim-denton2015}.

\begin{figure}
\includegraphics[width=0.9\columnwidth]{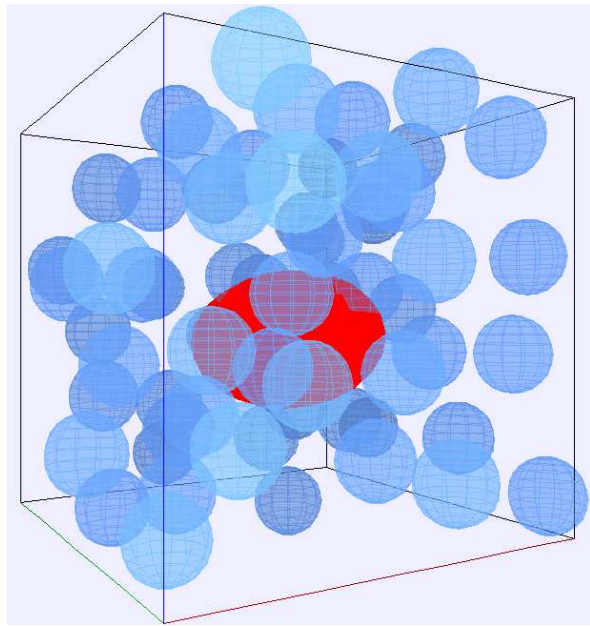}
\caption{
Snapshot from simulation of polymer-nanoparticle mixture for 
computing shape distribution of ellipsoidal polymer immersed 
in a concentrated dispersion of nanospheres.
}\label{snapshots}
\end{figure}

In the second series of simulations, using the eigenvalue distributions of an uncrowded
polymer [Eqs.~(\ref{Plambda}) and (\ref{Eurich-Maass})], we computed the shape distributions 
of polymers immersed in bulk dispersions of nanospheres of various concentrations
(Fig.~\ref{snapshots}).  
For both polymers and nanospheres, the tolerance for trial displacements 
was fixed at 0.1 $\sigma_n$.
Extending our preliminary study of crowding~\cite{lim-denton2014}, 
we implemented an exact polymer-nanosphere overlap algorithm~\cite{heckbert1994}
and used a more realistic 
penetration free energy $\varepsilon$ [Eqs.~(\ref{epsilon}) and (\ref{volume-ellipsoid})].
Since $\varepsilon$ depends, at each nanosphere volume fraction $\phi_n$, on the 
average volume of the crowded polymer $v_p(\phi_n)$, which itself depends on $\varepsilon$, 
we input the $v_p(\phi_n)$ predicted by free-volume theory and subsequently
checked for self-consistency.  

In each of the five independent runs, we accumulated configurational data over $10^7$ steps,
following an equilibration stage of $5\times 10^4$ MC steps, and calculated the polymer 
gyration tensor eigenvalue distributions, asphericity, and rms radius of gyration
by averaging over $10^4$ independent configurations, spaced by intervals of $10^3$ steps 
to minimize correlations.
We typically chose $N_n=216$, but performed runs with up to 1728 nanospheres 
to ensure statistical independence of system size.
In the process, we confirmed that our previous approximation~\cite{lim-denton2014} 
for the shape of the intersection region as a ``stretched" ellipsoid -- an ellipsoid 
whose principal radii are lengthened by $R_n$ -- is quite reasonable 
in the protein limit and gives qualitatively consistent results.
Results reported below are compared with predictions of free-volume theory.  
To our knowledge, no corresponding data from simulations of explicit polymer models 
are yet available for direct comparison.

\begin{figure}
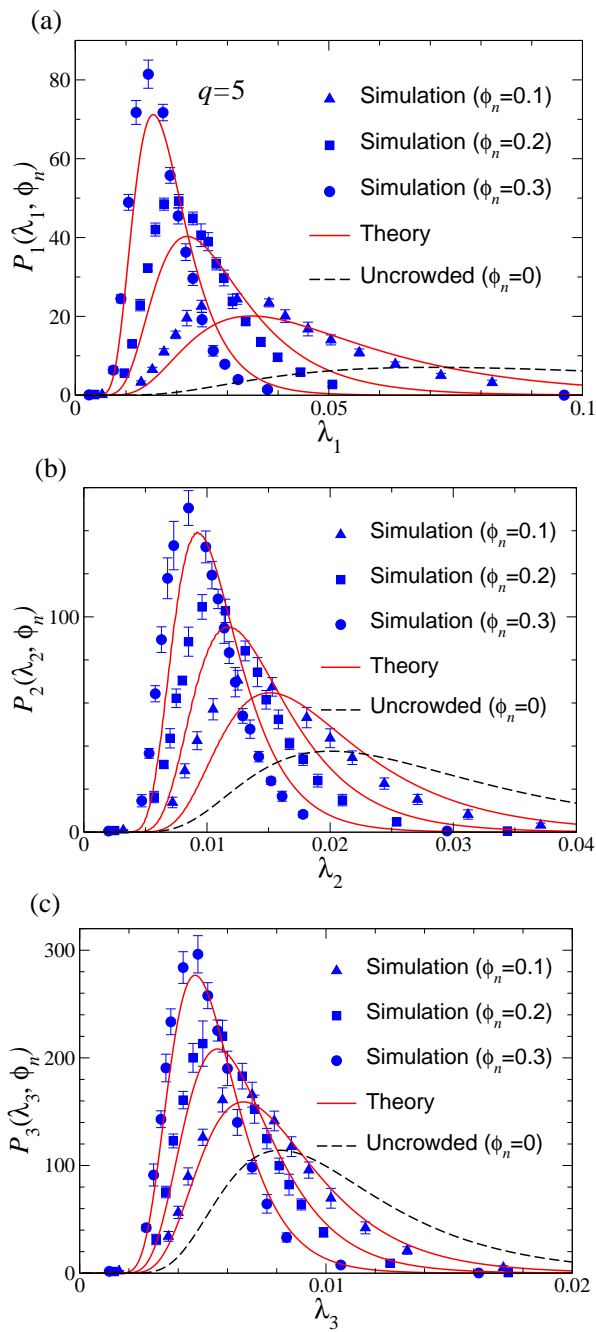

\includegraphics[width=0.9\columnwidth]{fig4a.eps} 
\includegraphics[width=0.9\columnwidth]{fig4b.eps} 
\includegraphics[width=0.9\columnwidth]{fig4c.eps} 
\vspace*{-0.2cm}
\caption{
Probability distributions for the eigenvalues of the gyration tensor of a crowded
polymer, modeled as an ideal, freely-jointed chain:  
(a) $\lambda_1$, (b) $\lambda_2$, (c) $\lambda_3$.
Our simulation data (symbols) are compared with predictions of free-volume theory 
(solid curves) for a single ellipsoidal polymer, with uncrowded rms radius of gyration 
equal to five times the nanoparticle radius ($q=5$), amidst $N_n=216$ hard nanospheres 
with volume fraction $\phi_n=0.1$ (triangles), 0.2 (squares), and 0.3 (circles).
Dashed curves show uncrowded distributions ($\phi_n=0$) from
Eqs.~(\ref{Plambda}) and (\ref{Eurich-Maass}).
}\label{fig-Plambda-q5}
\end{figure}
\begin{figure}
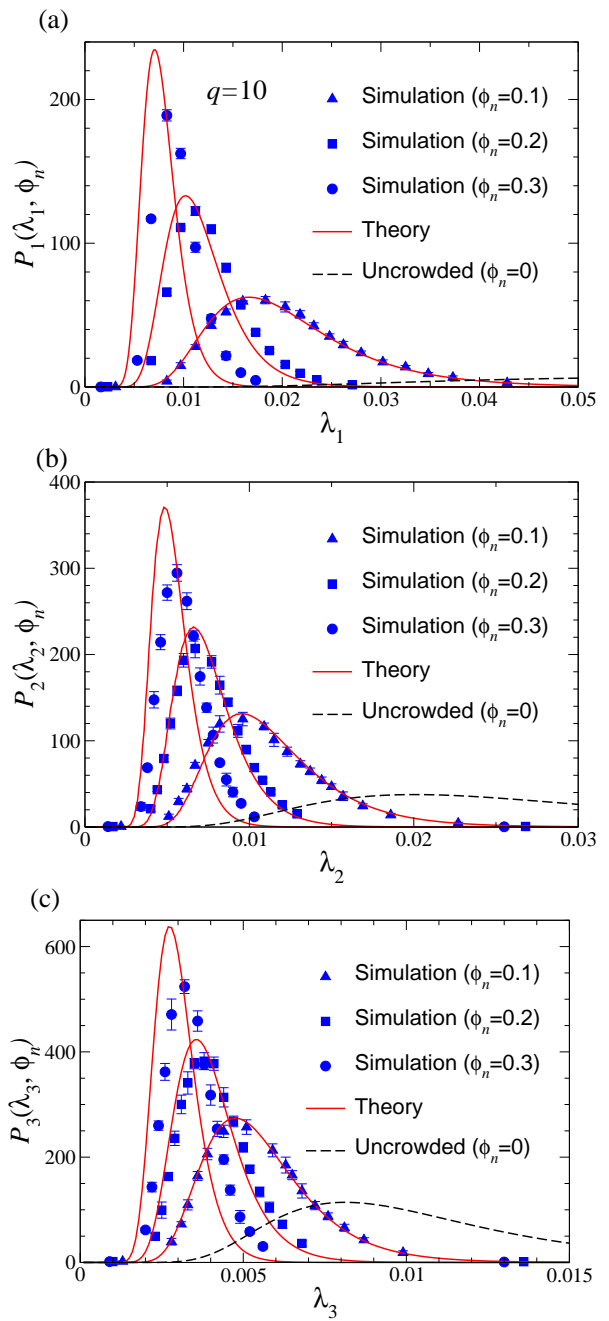

\includegraphics[width=0.9\columnwidth]{fig5a.eps} 
\includegraphics[width=0.9\columnwidth]{fig5b.eps} 
\includegraphics[width=0.9\columnwidth]{fig5c.eps} 
\vspace*{-0.2cm}
\caption{
Same as Fig.~\ref{fig-Plambda-q5}, but for larger polymer-to-nanosphere size ratio ($q=10$).
Notice the changes in scale.
\vspace*{-0.5cm}
}\label{fig-Plambda-q10}
\end{figure}

For a polymer immersed in a nanosphere dispersion, with uncrowded rms radius of gyration 
equal to five times the nanosphere radius ($q=5$), Fig.~\ref{fig-Plambda-q5} presents 
the probability distributions for the eigenvalues of the gyration tensor, representing 
the distribution of shapes of the ellipsoid that best fits the polymer coil.
A comparison of the scales of the distributions for the three eigenvalues reveals
that the typical shape of the polymer is that of an elongated, flattened ellipsoid.
With increasing crowding, i.e., nanosphere concentration, all three eigenvalue 
distributions steadily shift toward smaller ranges, reflecting contraction of the 
polymer along all three principle axes.  
Also shown in Fig.~\ref{fig-Plambda-q5} are predictions of free-volume theory 
(Sec.~\ref{theory}).  At lower nanosphere concentrations, simulation and theory 
agree closely.  With increasing crowding, however, small deviations emerge,
especially notable for the largest eigenvalue $\lambda_1$.  

Figure~\ref{fig-Plambda-q10} shows the probability distributions for a polymer 
doubled in size ($q=10$).  
Compared with the smaller polymer, for the same nanosphere concentration, the 
shifts in the eigenvalue distributions are significantly larger relative to the 
uncrowded distributions.  As discussed in ref.~\cite{lim-denton2014}, this trend 
can be understood by noting that the average overlap free energy increases roughly
with the square of the size ratio.  Also apparent is that the free-volume theory
is less accurate for this larger size ratio, somewhat overpredicting the
polymer compression, especially at the highest volume fraction.

\begin{figure}[t!]
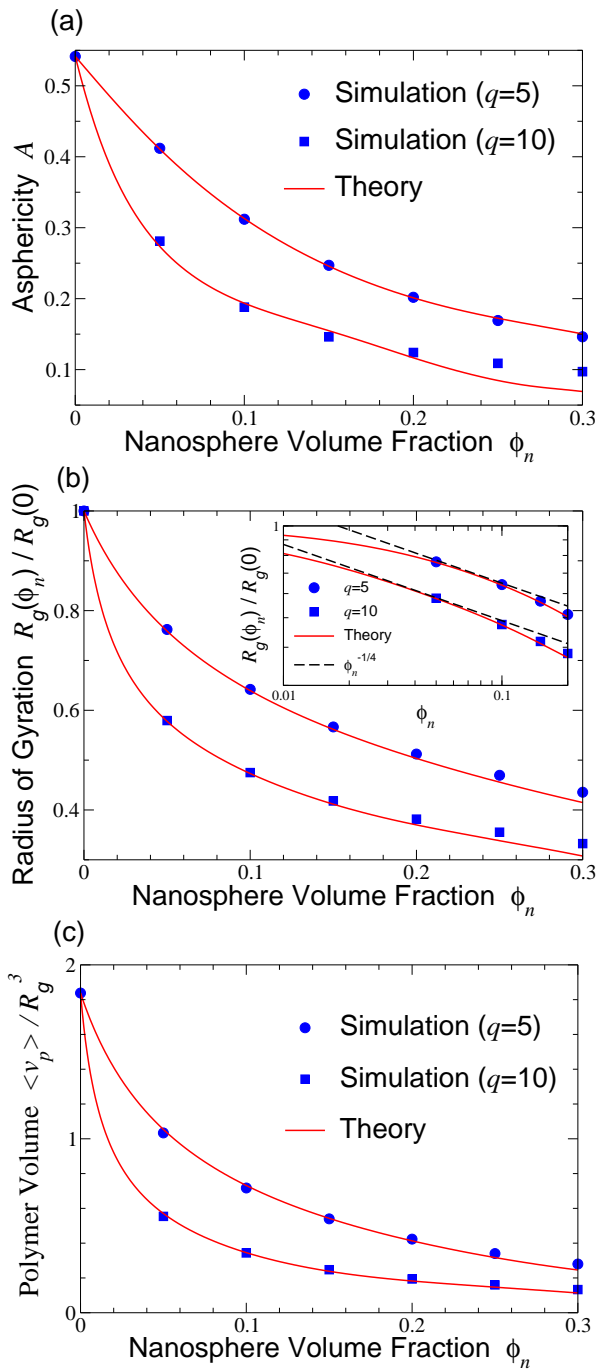

\includegraphics[width=0.9\columnwidth]{fig6a.eps}
\includegraphics[width=0.9\columnwidth]{fig6b.eps}
\includegraphics[width=0.9\columnwidth]{fig6c.eps}
\vspace*{-0.2cm}
\caption{
(a) Asphericity [Eq.~(\ref{asphericity})], (b) rms radius of gyration [Eq.~(\ref{gyration-ratio})],
and (c) average volume [Eq.~(\ref{volume-ellipsoid})] of a fluctuating ellipsoidal polymer 
vs.~nanosphere volume fraction $\phi_n$.  
Our simulation data are shown for uncrowded polymer-to-nanosphere size ratio $q=5$ (circles) 
and $q=10$ (squares) and compared with predictions of free-volume theory (curves).  
Error bars are smaller than symbols.
As crowding increases, the polymer becomes more compact (smaller and less aspherical).
Inset to panel (b): Comparison with scaling prediction~\cite{odijk2000}, $R_g\sim\phi_n^{-1/4}$,
on a log-log scale.
}\label{fig-shape}
\end{figure}

From the eigenvalue distributions, we computed the asphericity [Eq.~(\ref{asphericity})], 
rms radius of gyration [Eq.~(\ref{gyration-ratio})], and average volume 
[Eq.~(\ref{volume-ellipsoid})] of a crowded polymer over a range of nanosphere concentrations.  
As Fig.~\ref{fig-shape} demonstrates, a polymer responds to 
progressive crowding not only by contracting, but also by becoming more spherical, 
as reflected by the decrease in $A$, $R_g$, and $v_p$ with increasing $\phi_n$.  
These trends are amplified upon increasing the size ratio from $q=5$ to $q=10$, 
the larger polymer being relatively more compactified.  Moreover, the crowding effect 
is much stronger here, where we computed $\varepsilon$ from Eqs.~(\ref{epsilon}) and 
(\ref{volume-ellipsoid}), than in our previous study~\cite{lim-denton2014}, 
where we used lower penetration free energies.  Nevertheless, a model of spherical, 
compressible polymers gives yet greater contraction~\cite{lu-denton2011,lim-denton2014}, 
which is attributable to the lack of freedom of a spherical polymer to distort its shape.

Figure~\ref{fig-shape} also shows a comparison of our simulation data with predictions 
of free-volume theory.  As with the eigenvalue distributions, the theory faithfully 
captures the trends in shape, size, and volume.  The inset to panel (b) shows a 
further comparison with the scaling prediction of Odijk~\cite{odijk2000} for the
radius of gyration, $R_g\sim\phi_n^{-1/4}$, which is seen to be fairly consistent 
with our data over a significant range of nanosphere concentrations.  Here the 
close agreement between the input (theory) and output (simulation) values of 
$v_p$ serves as a consistency check on our approximation for the penetration free energy.
In the coarse-grained polymer model, for the size ratios studied here, 
neither simulation nor theory indicates a sudden collapse of an ideal coil 
up to volume fractions $\phi_n\simeq 0.4$.

\section{Conclusions}\label{conclusions}
In summary, to investigate influences of polymer shape and penetrability on mixtures of 
hard nanospheres and nonadsorbing homopolymers, modeled as penetrable ellipsoids with fluctuating shapes, 
we developed a Monte Carlo simulation method based on polymer insertion and geometric overlap 
algorithms.  We applied our method to compute depletion-induced potentials of mean force between 
nanospheres and crowding-induced shape deformations of ideal polymers in the protein limit.
Our simulation data for pair interactions are in good agreement with predictions of 
polymer field theories, further validating the ellipsoidal polymer model and demonstrating 
the importance of polymer shape fluctuations and penetrability for depletion interactions.
Our results for shape distributions of crowded polymers, including asphericity and 
rms radius of gyration, agree closely with predictions of free-volume theory, 
differing quantitatively only in highly concentrated nanosphere dispersions.  
Extending our previous study~\cite{lim-denton2014}, we consistently incorporated the
dependence of the penetration free energy on the polymer shape and nanoparticle concentration.
Furthermore, our predictions for polymer shape deformations can be tested against 
molecular simulations or density-functional theory calculations for explicit 
segmented-chain polymers, 
which may guide refinement of the step-function approximation assumed for 
the penetration energy profile.

The methods and results presented here lay a foundation for simulating more realistic
models of polymer-nanoparticle mixtures,
as well as models of polymers in quenched disordered media
~\cite{edwards-muthukumar1988,cates-ball1988,goldschmidt2000}.
Future work will focus on generalizing the 
model from ideal polymers to nonideal polymers in good solvents with 
excluded-volume interactions~\cite{bolhuis2003,doxastakis2004,doxastakis2005}
characterized by shape distributions of self-avoiding random walks~\cite{sciutto1996,schaefer1999},
analyzing crowding of 
real biopolymers~\cite{goldenberg2003,dima2004,cheung2005,denesyuk2011,denesyuk-thirumalai2013a}
(e.g., specific proteins and RNA), and exploring influences of polymer shape fluctuations
on thermodynamic stability and phase behavior (e.g., demixing) of bulk polymer-nanoparticle mixtures.

\acknowledgments
This work was supported by the National Science Foundation (Grant No.~DMR-1106331).
We thank Jan Forsman for helpful correspondence and for sharing the data from 
ref.~\cite{forsman2014} plotted in Fig.~\ref{fig-pmf-theory}.





%

\end{document}